\documentclass[12pt]{article}
\usepackage[pdftex]{graphicx} 
\usepackage{epsfig}
\usepackage{comment}
\usepackage{latexsym}
\usepackage{color}

\newcommand{\mysquare}[0]{\raise-.2ex\hbox{{\Large$\Box$}}}
\def\lsim{\mathrel{\rlap {\raise.5ex\hbox{$ < $}}
{\lower.5ex\hbox{$\sim$}}}}
\def\gsim{\mathrel{\rlap {\raise.5ex\hbox{$ > $}}
{\lower.5ex\hbox{$\sim$}}}} \topmargin -1.5cm \textheight=22.5cm \textwidth=16.5cm
\catcode`\@=11
\newcount\hour
\newcount\minute
\newtoks\amorpm
\hour=\time\divide\hour by60 \minute=\time{\multiply\hour by60
\global\advance\minute by-\hour}
\edef\standardtime{{\ifnum\hour<12 \global\amorpm={am}%
        \else\global\amorpm={pm}\advance\hour by-12 \fi
        \ifnum\hour=0 \hour=12 \fi
        \number\hour:\ifnum\minute<10 0\fi\number\minute\the\amorpm}}
\edef\militarytime{\number\hour:\ifnum\minute<10 0\fi\number\minute}
\def\draftlabel#1{{\@bsphack\if@filesw {\let\thepage\relax
   \xdef\@gtempa{\write\@auxout{\string
      \newlabel{#1}{{\@currentlabel}{\thepage}}}}}\@gtempa
   \if@nobreak \ifvmode\nobreak\fi\fi\fi\@esphack}
        \gdef\@eqnlabel{#1}}
\def\@eqnlabel{}
\def\@vacuum{}
\def\draftmarginnote#1{\marginpar{\raggedright\scriptsize\tt#1}}
\def\draft{\oddsidemargin -.2truein
        \def\@oddfoot{\sl preliminary draft \hfil
        \rm\thepage\hfil\sl\today\quad\militarytime}
        \let\@evenfoot\@oddfoot \overfullrule 3pt
        \let\label=\draftlabel
        \let\marginnote=\draftmarginnote
   \def\@eqnnum{(\theequation)\rlap{\k

 ern\marginparsep\tt\@eqnlabel}%
\global\let\@eqnlabel\@vacuum}  }

\newcommand{\be}[0]{\begin{equation}}
\newcommand{\ee}[0]{\end{equation}}
\newcommand{\ba}[0]{\begin{eqnarray}}
\newcommand{\ea}[0]{\end{eqnarray}}
\newcommand{\dis}{\displaystyle}

\def\bs{\begin{subequations}}
\def\es{\end{subequations}}

\def\thebibliography#1{%
\vskip 0.5cm \centerline{\bf \Large References}
\list{%
[\arabic{enumi}]}{\settowidth\labelwidth{[#1]} \leftmargin\labelwidth
\advance\leftmargin\labelsep
\usecounter{enumi}}
\def\newblock{\hskip .11em plus .33em minus .07em}
\sloppy\clubpenalty4000\widowpenalty4000 \sfcode`\.=1000\relax}

\renewcommand{\theequation}{\arabic{section}.\arabic{equation}}

\renewcommand{\section}{\setcounter{equation}{0}\@startsection
{section}{1}{0mm}{-\baselineskip}{0.5\baselineskip} {\normalfont\Large\bfseries}}

\renewcommand{\subsection}{\@startsection
{subsection}{2}{0mm}{-\baselineskip}{0.5\baselineskip} {\normalfont\large\bfseries}}

\renewcommand{\subsubsection}{\@startsection
{subsubsection}{3}{0mm}{-\baselineskip}{0.5\baselineskip}
{\normalfont\normalsize\slshape}}

\usepackage{amssymb}
\usepackage{amsthm}
\usepackage{amsmath}
\usepackage{amssymb,amsfonts}
\usepackage{graphicx}
\usepackage{cite}

\newcommand{\Z}{\mathbb{Z}}

\renewcommand{\S}{{\cal S}}
\newcommand{\N}{{\cal N}}

\newcommand{\abs}{|}
\newcommand{\where}{\mbox{where}}

\renewcommand{\and}{\mbox{and}}

\newcommand{\F}{{\cal F}}

\newcommand{\Th}[2]{\theta\left[^{#1}_{#2}\right]}

\begin{document}
\begin{titlepage}
\begin{flushright}
CPHT--RR077.0810, 
LPTENS--10/31, 
September 2010 
\end{flushright}

\vspace{0.2cm}

\begin{centering}
{\bf\Large Non-Singular String Cosmology\\ 
\vskip 0.2cm
in a 2d Hybrid Model}
\\

\vspace{0.9cm}
 {\large I. Florakis$^{1}$, C.~Kounnas$^{1}$, H. Partouche$^{2}$ and N. Toumbas$^{3}$  \\
 }

\vspace{0.9cm}

$^1$ Laboratoire de Physique Th\'eorique,
Ecole Normale Sup\'erieure,$^\dagger$ \\
24 rue Lhomond, F--75231 Paris cedex 05, France\\
{\em  florakis@lpt.ens.fr,~kounnas@lpt.ens.fr,~}
\vspace{2mm}

$^2$ Centre de Physique Th\'eorique, Ecole Polytechnique,$^\ddag$\\
F--91128 Palaiseau cedex, France\\
{\em herve.partouche@cpht.polytechnique.fr}

\vspace{2mm}

$^3$ Department of Physics,
University of Cyprus \\
Nicosia 1678, Cyprus\\
{\em nick@ucy.ac.cy}

 \vspace{0.8cm}

{\bf\Large Abstract}\\
\vspace{0.3cm}
The existence of non-singular string cosmologies is established in a class of two-dimensional supersymmetric Hybrid models at finite temperature. The left-moving sector of the Hybrid models gives rise to $16$ real $({\cal N}_4=4)$ spacetime supercharges as in the usual 
superstring models. The right-moving sector is non-supersymmetric at the massless level, but is characterized by \emph{MSDS} symmetry, which ensures boson/fermion degeneracy of the right-moving massive levels. Finite temperature configurations, which are free of Hagedorn instabilities, are constructed in the presence of non-trivial ``gravito-magnetic'' fluxes. These fluxes inject non-trivial winding charge into the thermal vacuum and restore the thermal $T$-duality symmetry associated with the Euclidean time circle. Thanks to the unbroken right-moving \emph{MSDS} symmetry, the one-loop string partition function is exactly calculable beyond any $\alpha^{\prime}$-approximation. At the self-dual point new massless thermal states appear, sourcing localized spacelike branes, which can be used to connect a contracting thermal Universe to an expanding one. The resulting bouncing cosmology is free of any curvature singularities and the string coupling remains perturbative throughout the cosmological evolution.    
\end{centering}


\noindent 
\vspace{3pt} \vfill \hrule width 6.7cm \vskip.1mm{\small \small \small \noindent \\
$^\dagger$\ Unit{\'e} mixte  du CNRS et de l'Ecole Normale Sup{\'e}rieure associ\'ee \`a
l'Universit\'e Pierre et Marie Curie (Paris
6), UMR 8549.\\
$^\ddag$\ Unit{\'e} mixte du CNRS et de l'Ecole Polytechnique,
UMR 7644.}
\end{titlepage}
\newpage
\setcounter{footnote}{0}
\renewcommand{\thefootnote}{\arabic{footnote}}
 \setlength{\baselineskip}{.7cm} \setlength{\parskip}{.2cm}

\setcounter{section}{0}


\section{Introduction}

One of the most acute open problems of Modern Cosmology is that of the initial singularity. 
Assuming that the matter content of the Universe satisfies the weak or the null energy condition, 
and extrapolating the cosmological evolution arbitrarily back in time by using the equations of quantum field theory 
and Einstein's theory of general relativity, 
one is driven to a curvature singularity, the ``Big Bang'', signifying a break-down of our description of the underlying physics. 
Moreover, the singularity theorems \cite{HE} strongly suggest that it is unlikely to overcome this problem
within the framework of quantum field theory and general relativity.

In string theory, however, new degrees of freedom arise and various
non-geometrical phenomena take place, opening up new possibilities to address the initial singularity problem \cite{CosmoPhenomena, KL,CosmoTopologyChange,GV,R,LO,BFM,massivesusy,KW}.
Indeed, the full set of stringy degrees of freedom cannot be incorporated 
within a single effective field theory
description. Rather, there exist distinct effective descriptions valid in 
specific restricted regions of the moduli space. Typically in such regions, most stringy states carry large masses of the order of the string scale or higher. At low enough energies, these can be integrated out giving rise to an effective field
theory in terms of a finite number of light fields. However, as the moduli fields 
vary adiabatically, we encounter points where new states become relevant, 
forcing us to switch from the initial description into another, using the duality symmetries of
string theory. An apparently singular geometry may then be mapped into a non-singular, smooth geometry in the dual description. 
Furthermore, around the transition points a conventional field theory description may be absent, 
as the full string theory dynamics becomes relevant. For comprehensive  
discussions, see e.g. \cite{CosmoTopologyChange}. 

In a cosmological context, where a number of moduli fields and the background metric acquire time dependence,  
we expect the various dual descriptions to be connected dynamically via non-trivial phase transitions \cite{CosmoPhenomena,AW,RK,AKADK,CosmoTopologyChange,BFM,massivesusy}. 
The cosmological evolution will be marked by transition 
eras, where new stringy states become light and dominate the dynamics. In general, the transitions 
among the various dual phases take place at finite values of the moduli, where the underlying target space background has finite size and/or curvature. Thus, one expects these transitions to occur before the appearance of cosmological
singularities. The hope is that, once the underlying string dynamics is properly taken into account, the various non-singular cosmological phases would be connected consistently, successfully covering the whole history of the cosmological evolution. Previous work on string dualities and cosmology includes \cite{CosmoPhenomena, KL,CosmoTopologyChange,GV,R,LO,BFM,massivesusy,KW}. 

The greatest obstacle towards a concrete realization of this attractive cosmological scenario is the appearance of 
Hagedorn instabilities 
in string theory at finite temperature and in the presence of other supersymmetry-breaking sources. 
Due to the exponential growth in the number of single-particle states as a function of mass, 
the partition function
of superstrings at finite temperature diverges once the temperature exceeds the Hagedorn temperature ($T_H \sim M_s$). 
This divergence, however, is not a pathology of string theory. Rather, it signals a phase transition towards a new thermal vacuum \cite{KoganS,AW,AKADK,BR}. 
In the Euclidean description of the thermal system, where the time direction is compactified on a circle of 
radius $R_0$, $T =1/(2\pi R_0)$, certain stringy states winding the Euclidean time circle become tachyonic 
precisely when the thermal modulus $R_0$ exceeds its Hagedorn value. 
Thus, the Hagedorn divergence can be interpreted as an IR-instability of the underlying Euclidean background and
the phase transition is driven by tachyon condensation. The crucial point is that, in the presence of such condensates, the 
thermal vacuum acquires non-trivial winding charge\cite{AW,AKADK}. The question that we would like to address is whether it is possible to identify stable thermal vacua, characterized by non-zero winding charge, whose dynamics could be quantitatively treated by utilizing e.g. the powerful 2d worldsheet CFT techniques of (weakly coupled) string theory.    

Considerable progress towards this direction has been achieved in the context of $(4,0)$-supersymmetric type II vacua in arbitrary dimension, where
it was found that the introduction into the thermal system of certain discrete ``gravito-magnetic'' fluxes, 
associated with the
graviphoton ($G_{I0}$) and the axial vector ($B_{I0}$) backgrounds, lifts the tachyonic instabilities \cite{akpt,FKT}. 
Essentially,
these fluxes inject non-trivial winding charge into the thermal vacuum, and restore the stringy $T$-duality symmetry
$R_0 \to R_c^2/R_0$ of the finite-temperature system\footnote{In the absence of the gravito-magnetic fluxes
the type II thermal partition function does not enjoy thermal duality symmetry. In the heterotic case, however, this
duality symmetry is present.}. 
The self-dual point occurs at the fermionic point $R_c = 1/\sqrt{2}$, where new 
massless thermal states appear. These states correspond to vortices carrying
non-trivial momentum and winding charges along the Euclidean time circle, and lead to an enhancement of the local $U(1)_L$ gauge symmetry 
to a non-Abelian $[SU(2)_L]_{k=2}$ gauge symmetry.
The one-loop string partition function is finite for all values of the thermal
modulus $R_0$ and reduces to the conventional thermal partition function in the large $R_0$ limit. 

In this paper we will study the induced cosmological evolution associated with such thermal
configurations in a class of two dimensional $(4,0)$-vacua, the so called Hybrid vacua, introduced
in \cite{FKT}. These are very special vacua: the right-moving supersymmetries are broken spontaneously at
the string level, but are replaced by the recently-discovered Massive (boson/fermion) Spectrum Degeneracy
Symmetries (\emph{MSDS}) of \cite{massivesusy,reducedMSDS}.
The presence of particular discrete ``gravito-magnetic''
fluxes renders the finite temperature configurations free of tachyonic/Hagedorn divergences and, in addition, the thermal vacuum is characterized by unbroken right-moving \emph{MSDS} symmetry.
As a result of the latter symmetry, the one-loop string partition function can be computed exactly in $\alpha^{\prime}$,
with the contributions of the massive bosonic and fermionic towers of string states\footnote{Notice that, at the fermionic point, one may treat lattice states as fermionic oscillator states in a uniform manner.} exactly cancelling each other.
As the value of the thermal modulus $R_0$
is varied, the thermal system exhibits an interesting phase transition, 
occurring 
precisely at the extended symmetry point $R_0=1/\sqrt{2}$. At this point additional massless states appear,
driving the phase transition.
 The two phases are related by thermal duality symmetry $R_0 \to 1/(2R_0)$ (with the extended symmetry
 point being the self-dual point)\footnote{Various aspects of
thermal duality symmetry have been also discussed in \cite{Os,Chaud,DL}.}. In both
phases the temperature, the energy density and pressure are bounded. They achieve their maximal
critical values at the extended symmetry point. Thanks to the unbroken \emph{MSDS} symmetry, 
the equation of state characterizing each phase
is effectively that of massless thermal radiation in two dimensions.

The resulting cosmological evolution is non-singular, and it is marked by a phase transition between
the two dual thermal phases. We will show that the phase transition can be described in terms of
a space-like brane, corresponding to a localized pressure that is sourced by the additional massless
thermal modes that become relevant at the critical temperature. The role of the brane is to glue 
the two dual thermal phases in a consistent way. As we will see, both the dilaton and the
scale factor bounce across the brane. For the scale factor, the bounce corresponds to a smooth reversal
of contraction to expansion, with the Big Bang curvature singularity of general relativity being absent. 
There is a discontinuity in the first time derivative of the dilaton,
determined by the tension of the brane. This discontinuity is naturally resolved by the presence of extra massless thermal states
localized at the brane. The system remains weakly coupled throughout the cosmological evolution, with
the dilaton attaining its maximal value at the phase transition. This maximal value is determined
in terms of the brane tension, the critical temperature and the number of the thermally excited massless degrees
of freedom in the system. 

We argue that the gluing mechanism provided by such spacelike branes is
a fundamental dynamical phenomenon in string theory, capable of connecting consistently the space of ``momenta" with
the dual space of ``windings". Such branes will typically appear at the self-dual points, which are characterized by
enhanced gauge symmetry. The presence of additional marginal operators with non-trivial
momentum and winding charges is crucial in order for a dynamical transition between the two dual spaces to occur.

We would like to stress here that our stringy, singularity-free cosmological solution differs from various incarnations of the pre- Big Bang
scenario \cite{GV}, which require the application of a non-perturbative S-duality symmetry in order
to resolve the very early, pre- Big Bang phase. In fact, the occurrence of strong coupling dynamics 
is often another great obstacle in realizing the most general cosmological scenario outlined above. It is an open and interesting question to find branes that can dynamically glue cosmological phases that are related by S-duality symmetry. It is also interesting that the two-dimensional cosmological Hybrid models
are amenable to perturbative treatment throughout their entire history.      

The plan of this paper is as follows:

In section \ref{40} we summarize the main features concerning tachyon-free, thermal configurations
of $(4,0)$-vacua at finite temperature. In section \ref{TH} we focus on the two dimensional
Hybrid models and analyze in detail the resulting phase structure of the thermal system.
We proceed in section \ref{Cosmo} to study the backreaction of the thermal Hybrid state on the 
initially flat metric and dilaton background and exhibit the resulting non-singular cosmology.
In section \ref{General}, we present more general non-singular cosmological
solutions associated to distributions of spacelike branes.
We end up with our conclusions and directions for future research.

\section{Thermal (4,0)-type II string vacua}
\label{40} 
Before describing the two dimensional Hybrid models, 
we review
some general results concerning thermal states of 
$(4,0)$-type II models \cite{akpt,FKT}. 
Such models correspond to freely-acting asymmetric orbifold
compactifications, which preserve  
the $16$ real supersymmetries (${\cal{N}}_4=4$) 
arising from the left-moving
sector of the worldsheet. The right-moving supersymmetries are broken
spontaneously at the string level, via the coupling of an internal cycle, e.g. the $X^9$ direction compactified on a circle,
to the right-moving spacetime fermion number $F_R\equiv \bar a$. 
At special values of the internal compactification moduli, we 
encounter extended symmetry points with a non-Abelian enhancement 
of the local gauge symmetry group \cite{akpt,FKT}.
In the following discussion, we take
the values of the internal moduli to be frozen at such 
an extended symmetry point
and the string coupling to be sufficiently weak.    

The finite temperature partition function can be 
computed via a Euclidean path integral on
the $S^1(R_0) \times {\cal{M}}$ manifold, where the Euclidean time direction is compactified on a circle of radius $R_0$. 
Here ${\cal{M}}$ denotes the 
spatial manifold. At least one spatial direction is taken to be very large 
(or non-compact). 
The conventional thermal
ensemble corresponds to a freely acting orbifold, 
obtained by modding out with      
$(-1)^F\delta_0$, where $\delta_0$ is an order-two shift 
along the Euclidean time circle. 
The coupling to
the spacetime fermion number $F\equiv a+\bar a$ 
is dictated by the spin-statistics connection.
The resulting thermal system
develops instabilities, the
Hagedorn instabilities, when $R_0 < R_H=\sqrt{2}$, 
where certain stringy states   
winding the Euclidean time circle become tachyonic \cite{KoganS,AW,RK}.
These instabilities signal the onset of a non-trivial phase
transition around the Hagedorn temperature, 
see e.g. \cite{AW,AKADK,BR}. 
Generically, one expects the
tachyon condensates involved in the transition to lead to large 
backreaction, 
driving the system outside the perturbative domain. A precise quantitative
analysis of the dynamics is still lacking.
At the very least, we can identify 
a crucial property of the would-be stable high-temperature phase: 
since certain winding modes acquire non trivial expectation values, 
$\langle O_n\rangle \,\ne\, 0$, the thermal vacuum must be 
characterized by non-zero winding charge. 

In this paper, we will consider tachyon-free thermal configurations 
where, in addition to the temperature deformation, we
turn on certain discrete ``gravito-magnetic'' fluxes
associated with the graviphoton ($G_{I0}$)
and the axial vector ($B_{I0}$) gauge fields. These are geometrical fluxes 
threading the Euclidean time circle together with
the internal cycle responsible for the breaking of 
the right-moving supersymmetries \cite{akpt,FKT}. 
Effectively, they describe
gauge field ``condensates'' which have locally vanishing field strength 
but a non-zero value of the Wilson
line around the Euclidean time circle\footnote{See e.g. \cite{thermalgauge} for 
discussions of such configurations in the context of local gauge theories at
finite temperature and \cite{Chaud,G} in the context of string theory.}. They correspond to non-trivial,
topological vacuum parameters, defining (at least perturbatively) 
super-selection sectors of the thermal system. 
The presence of such fluxes modifies the thermal frequencies and the 
contribution to the free energy
of various states charged 
under the graviphoton and the axial vector gauge fields. 
As the would-be winding tachyons are
charged under these gauge fields, 
for large enough values of the ``gravito-magnetic fluxes'', 
the tachyonic
instabilities are lifted \cite{akpt,FKT}. 

As an illustrative example, consider the case where the
$X^9$-cycle factorizes from the rest of the internal compact manifold \cite{akpt}. In the presence of non-zero $G_{90}/R_9^2\equiv G$ and $B_{90}/R_9^2\equiv B$ background fields and
finite temperature, 
the contribution to the one-loop partition function of the $(X^0,\, X^9)$ sub-lattice is given by
$$
\frac{R_0 \, R_9}{\tau_2} \sum_{\tilde m , n}  
e^{-\frac{\pi}{\tau_2}\, \left[ R_0^2 |\tilde m ^0 + \tau n^0 |^2 + 
R_9^2 |\tilde m^9 + G \tilde m^0 +\tau (n^9 + G n^0 )|^2 \right]} \, 
\times e^{2 i \pi B (\tilde m^9 n^0 - \tilde m^0 n^9 )}
\nonumber
$$
\be
 \times ~(-1)^{\tilde m^0 (a+\bar a ) + n^0 (b+\bar b)}\, (-1)^{\tilde m^9 \bar a + n^9 \bar b + \tilde m^9 n^9} \, .
\ee
The Euclidean-time  $X^0$-cycle couples to the total spacetime fermion number $F$ and
the internal spatial $X^9$-cycle couples to the right-moving fermion number $F_R$.
The conventional finite-temperature system (suffering from the well-known Hagedorn pathologies) corresponds to $G=B=0$, where
the $X^0$-lattice coupling to the total fermion number factorizes. The deformed {\it tachyon-free} configuration
arises at the point $G=2B=1$ \cite{akpt}. This point is very special since it leads to an 
asymmetric factorization of the original lattice into two parts, one of which couples only to the left-moving $R$-symmetry charges, while the second one couples only to the right-moving $R$-symmetries. To see this, shift $\tilde m^9 \to \tilde m^9 -\tilde m^0$ and 
$n^9 \to n^9 -n^0$ to define the frame in which the $(2,2)$-lattice 
factorizes into two $(1,1)$-lattices, asymmetrically coupled to the left- and right-moving spacetime
fermion numbers $F_L \equiv a$, $F_R \equiv \bar a$, respectively:    
\be
\label{lattice09}
\frac{R_0}{\sqrt{\tau_2}}\sum_{\tilde m^0,n^0} 
e^{-\frac{\pi R_0^2}{\tau_2}|\tilde m^0 +n^0\tau|^2} (-1)^{\tilde m^0 a+n^0b+\tilde m^0 n^0}
\times
\frac{R_9}{\sqrt{\tau_2}}\sum_{\tilde m^9,n^9} 
e^{-\frac{\pi R_9^2}{\tau_2}|\tilde m^9 +n^9\tau|^2} (-1)^{\tilde m^9 \bar a+n^9\bar b+\tilde m^9 n^9}.
\ee
In this representation, the $X^0$ circle couples to the left-moving spacetime fermion number 
$F_L\equiv a$, and so the configuration can be described
as a freely-acting asymmetric orbifold $(-1)^{F_L}\delta_0$ 
of the initially supersymmetric $(4,0)$ model.

In \cite{FKT}, general classes of $(4,0)$-vacua were studied, including cases where
the $X^9$-cycle does
not factorize from the rest of the compact spatial manifold. 
It was found that the point $G=2B=1$ 
leads to the \emph{unique} tachyon-free, thermal configuration associated 
with any such initially supersymmetric $(4,0)$-vacuum. Moreover, 
it was established
that there exists a frame in which the full compactification lattice admits
an asymmetric factorization, where the $(1,1)$ sub-lattice associated to the Euclidean time cycle couples only to $F_L$, while the internal, 
spatial sub-lattice couples only to $F_R$:
\be\label{Factorization}
	\Gamma_{(d,d)}[^{a,~\bar{a}}_{b,~\bar{b}}]=\Gamma_{(1,1)}[^{a}_{b}](R_0)\otimes
\Gamma_{(d-1,d-1)}[^{\bar{a}}_{\bar{b}}](G_{IJ},B_{IJ})~.
\ee
This result generalizes (\ref{lattice09}). The thermal configurations
remain tachyon-free under all possible deformations of
the dynamical, transverse moduli associated with the internal
spatial sub-lattice, and of the thermal modulus $R_0$ \cite{FKT}.
Indeed when $G=2B=1$, the $O_8\bar O_8$ sector, which appears in the spectrum, 
necessarily carries 
non-trivial momentum and winding charges, 
and its lowest mass satisfies \cite{FKT}
\be
 m^2_{{ O} { \bar O}}~\ge ~\left( \frac{1}{2 R_0} -  R_0 \right)^2 ,
\ee
for any deformation of the transverse, internal spatial lattice. As a result the lowest-lying thermal states coming from the dangerous $O_8\bar{O}_8$-sector are at least massless (for details see \cite{FKT}).

Thus, the lowest mass in the $O_8\bar O_8$ sector is never tachyonic but becomes massless when $R_0={1 / \sqrt{2}}$ and for particular values of the transverse moduli in the internal spatial lattice, giving rise to enhanced symmetry points. The one-loop string partition function is finite for all values of the thermal modulus $R_0$, and it enjoys the $T$-duality symmetry: $R_0 \to 1/(2R_0)$. The self-dual fermionic point $R_0 =1 /\sqrt{2}$ corresponds to the extended symmetry
point.  

Let us focus on a special point in moduli where the local Abelian gauge symmetry associated
to the $X^0$ and $X^9$ cycles is extended to:
\be
U(1)_L^2 \times U(1)_R^2 \times \cdots \to \left([SU(2)_L]_{k=2} \times U(1)_L \right) 
\times \left( U(1)_R \times [SU(2)_R]_{k=2}\right)\times \cdots~.
\ee  
The first factor and the second factor in each parenthesis on the r.h.s. correspond to the $X^0,\, X^9$-cycles respectively. This enhanced gauge symmetry allows us to re-interpret the deformation 
in terms of the ``gravito-magnetic'' fluxes as
condensates of vortices 
carrying non-trivial momentum and winding charges 
along the Euclidean-time and the $X^9$-cycles. Indeed, the marginal deformation switching on the ``gravito-magnetic" fluxes is:
\be
	R_9^2 (G +2B)\left( \partial X^0\bar\partial X^9\right)+ 	R_9^2 (G -2B) \left( \partial X^9\bar\partial X^0\right).
	\ee
When $G=2B=1$, the second term vanishes. At the enhanced symmetry point, we are free to perform an $SU(2)_L \times SU(2)_R$ rotation\footnote{The three $[SU(2)_L]_{k=2}$-currents are $\partial X^0$, $\psi^0 \cos X^0_L$ and $\psi^0 \sin X^0_L$. At the enhanced symmetry point one may choose any one of the three. Notice that the $SU(2)$-symmetry currents under consideration also utilize the fermionic worldsheet super-partner $\psi^0$ \cite{RK}.  } in order to re-express the current-current deformation as: 
\be
\partial X^0 \bar\partial X^9 
\to \left(\psi^0 \cos X^0_L\right) \times \left(\bar \psi^9 \cos X^9_R\right) ~, 
\ee
where $\psi^0$ ($\bar \psi^9$) is the worldsheet super-partner of $X^0_L$
 ($X^9_R$). In other words, the deformation injects into the thermal vacuum non-trivial momentum
and winding
charges along the Euclidean-time and the $X^9$- cycles.  
The contribution to the free energy of the massive string states
is effectively regulated, resolving the would-be divergences. 
This observation illustrates how the Hagedorn
instabilities can be lifted once the Euclidean thermal vacuum acquires non-zero winding charge.

As we will see, the (Hagedorn-free) deformed thermal vacua 
exhibit interesting phase transitions. In the cosmological context, 
such tachyon-free thermal configurations will be
considered in full detail in a class of two-dimensional $(4,0)$-vacua, the so-called Hybrid 
models, introduced recently in \cite{FKT}.  
In this class of models, 
the modified free energy in each phase
satisfies all the usual thermodynamical properties as 
a function of the temperature. Namely, 
as the temperature increases, the free energy decreases 
monotonically up to the phase
transition and the specific heat is positive. At the phase transition, the temperature achieves a maximal critical value. As we will see,
the resulting phase structure is not only free of the Hagedorn instabilities, but also the
induced cosmology turns out to be free of the initial ``Big Bang'' 
curvature singularity.

\section{The tachyon-free thermal Hybrid model}
\label{TH}
In this section 
we study thermal configurations of the two-dimensional Hybrid models 
introduced in \cite{FKT}.
The Hybrid models describe very special $(4,0)$-supersymmetric 
vacua: 
the right-moving supersymmetries 
are broken spontaneously at the string level 
via the coupling of an internal cycle to $F_R$, 
and are replaced by the Massive Spectrum 
Degeneracy Symmetries (\emph{MSDS}) of
\cite{massivesusy,reducedMSDS}.
As in type II superstrings, 
we can define two versions of the model, 
Hybrid A and Hybrid B, 
depending on the relative chirality of the left- and right-moving 
Ramond sectors. 
In the maximally symmetric case,
the partition functions are given by
\be
 Z = {V_2\over (2\pi)^2}~\int_\F {d^2\tau\over
4({\rm Im}\tau)^{2}} 
~\left[\frac 12 \sum_{a,b=0,1} (-)^{a+b+\mu ab} {\Th ab^{4} \over \eta^{4}}\right]
\left[\frac 12 \sum_{\gamma,\delta=0,1} ~ {\theta[^{\gamma}_{\delta}]^{8} \over \eta^{8}}\right]
\left[\frac 12 \sum_{\bar a, \bar b=0,1} (-)^{\bar a+\bar b} ~ {\bar \theta[^{\bar a}_{\bar b}]^{12} \over \bar \eta^{12}}\right],
\ee
where $\mu=0$ in Hybrid B ($\mu=1$ in Hybrid A) and $V_2$ stands for the volume of the two large (Euclidean) directions $X^0, X^1$. 
In terms of $SO(n)$ left- and right-moving characters, the partition function takes the form 
\be
\label{2Z}
Z={V_2\over (2\pi)^2}~\int_\F {d^2\tau\over
4({\rm Im}\tau)^{2}}\, \left(\bar V_{24}-\bar S_{24}\right)\, {1 \over \eta^8}\, \Gamma_{E_8}(\tau)\left\{\!\!\begin{array}{l}\left(V_8 - S_8\right)\\\left(V_8 - C_8\right)\end{array}\right.,
\ee
where the upper (lower) entry refers to the Hybrid B (Hybrid A) case. 
Thus, the left-moving sectors are described in terms of the $SO(8)$-characters, as in conventional superstring models, 
and in terms of the chiral $E_8$-lattice. 
The right-moving sectors are described in terms of the \emph{MSDS} $SO(24)$-characters.
In particular, the right-moving characters satisfy the identity
\be
\bar V_{24}-\bar S_{24}=24, 
\ee
exhibiting the Massive Spectrum Degeneracy Symmetry of the right-moving sector.

The massless content of the Hybrid B (Hybrid A) model consists of $24 \times 8$ bosons and $24 \times 8$ fermions,  
arising from the $V_8~ \bar V_{24}$ and $S_8~ \bar V_{24}$ ($C_8~ \bar V_{24}$) sectors, respectively. There are no massless fermions 
from the right-moving R sector and the RR fields are massive. The
model can also be described in terms of a freely-acting asymmetric orbifold compactification of the type II superstring to two
dimensions. The relevant half-shifted $(8,8)$-lattice is given by \cite{FKT} 
\be
\label{CompactLattice}
\Gamma_{(8,8)}[^{\bar{a}}_{\bar{b}}]=\Gamma_{E_8}\times\bar\theta[^{\bar{a}}_{\bar{b}}]^8 \equiv \!\!\!
\sum\limits_{\tilde{m}^I,n^J;I,J=2,\dots,9} \!\!\!\frac{\sqrt{\det{G_{IJ}}}}{(\sqrt{\tau_2})^{8}}~  {e^{-\frac{\pi}{\tau_2}
(G+B)_{IJ}(\tilde{m}^I+\tau n^I) (\tilde{m}^J+\bar{\tau} n^J)}~}
~{e^{i\pi(\tilde{m}^9\bar a+ n^9\bar b+ \tilde{m}^9n^9)}},
\ee
where all internal radii are at the fermionic point\footnote{Note that in this representation the fermionic point does \emph{not} simply correspond to $R_i=1/\sqrt{2}$, as would be the case in simpler compactifications. The reason is the non-trivial coupling to $F_R$ which asymmetrically deforms the lattice. The specific values of the $(8,8)$-moduli corresponding to the fermionic/\emph{MSDS} point are given in \cite{FKT}.}. 
The modular covariant cocycle describes the coupling of 
a single internal cycle of the lattice, 
which we will henceforth refer to as the $X^9$-cycle, to $F_R$ 
and leads to the 
spontaneous breaking of the right-moving supersymmetries. 
At the \emph{MSDS} point the $G_{IJ}$ and $B_{IJ}$ tensors take very special values.
In fact, the \emph{MSDS} structure characterizing the right-moving sector gives rise to
an enhanced local non-Abelian gauge group: $U(1)_L^8 \times [SU(2)_R]_{k=2}^8$. A detailed study of
the classical moduli space of the \emph{MSDS} and Hybrid vacua, 
as well as the study of marginal deformations
connecting these vacua to four dimensional supersymmetric vacua, can be found in \cite{FKT}. 

At finite temperature, the backreaction on the initially flat metric and
dilaton background induces a cosmological evolution. 
As we saw in section \ref{40}, thermal tachyon-free         
configurations, able to bypass the Hagedorn instabilities, can be
constructed in terms of freely-acting asymmetric orbifolds along
the Euclidean time circle. In particular, these are obtained 
by modding out with $(-1)^{F_L}\delta_0$, where $\delta_0$ is a
$\Z_2$-shift along the Euclidean time circle, and
can be interpreted as discrete deformations 
of the conventional finite temperature
models. An important result is that they 
remain tachyon-free under arbitrary marginal deformations of the transverse 
dynamical moduli associated with the compact internal eight-manifold \cite{FKT}.
We will describe below some of their main features. 
Our notations concerning the wrapping numbers ${(\tilde m , n)}$, as well as  the the left- and right-moving momenta $(p_L,p_R)$
of the $\Gamma_{(1,1)}$-lattice are as follows:
$$
\Gamma_{(1,1)} (R_0) = ~\frac{R}{\sqrt{\tau_2}}\sum_{\tilde m , n} 
e^{-\pi \frac{R_0^2}{\tau_2} |\tilde m + n \tau |^2}
=~ \sum_{m,n} \Gamma_{m,n} \,,             
$$ 
\be
\label{lattice}
\Gamma_{m,n} = q^{\frac{1}{2} p_{ L}^2 }\, \bar q ^{\frac{1}{2} p_{ R}^2}\;,\quad   
q=e^{2\pi i \tau}\;,\quad \mbox{and}\quad      p_{ L,R} = {1 \over \sqrt{2}}\left(\frac{m}{R_0} \pm n R_0\right)\,.
\end{equation}

For sufficiently weak string coupling, the thermodynamic behavior of the system is captured
by the one-loop partition function. In the Hybrid B case, this is given by \cite{FKT, akpt}
\be
{Z \over V_1} = \int_{\cal F} \frac{d^2 \tau}{8\pi({\rm Im}\tau)^{3/2}}  
{\Gamma_{E_8}\over \eta^8}\sum_{m,n}
\left(V_8 \Gamma_{m,2n} + O_8 \Gamma_{m +\frac{1}{2},2n +1} - S_8  
\Gamma_{m +\frac{1}{2} , 2 n} -C_8 \Gamma_{m, 2n +1} \right) (\bar V_{24}- \bar S_{24}),
\label{parti}
\end{equation}
where $V_1=2\pi R_1$, whereas the Hybrid A case is obtained by the chirality exchange $S_8\rightarrow C_8$. 
The $(-1)^{F_L}\delta_0$ orbifold acts thermally 
on the left-movers and breaks the $(4,0)$-supersymmetries spontaneously. On the
other hand,
the right-moving \emph{MSDS} structure remains unbroken. 
In the odd-winding sector, the left-moving GSO projection is reversed, 
and so the
$O_8$- and $C_8$- sectors appear in the spectrum.
As the right-moving sector, however, 
begins at the massless level, 
the model remains tachyon-free for all values of the thermal modulus $R_0$ (as follows by level-matching and the properties of the $\Gamma_{m+\frac{1}{2},2n+1}$-lattice).
The Hybrid B and Hybrid A models are mapped into each other under the $T$-duality transformation $R_0 \to 1/(2R_0)$. 
The thermal quanta of Hybrid A are mapped into vortices, carrying non-trivial winding number, in the Hybrid B model and vice-versa.

As we already stressed, the one-loop partition function is finite
for all values of the thermal modulus $R_0$. In the Hybrid model, the partition function can be computed explicitly thanks 
to the \emph{MSDS} identity $\bar V_{24}- \bar S_{24}=24$. The result is
exact in $\alpha^{\prime}$, and it is given in terms of a 
surprisingly simple expression \cite{FKT}:
\be \label{partfinal}
{Z \over V_1}=24\times \left(R_0 + {1 \over 2R_0}\right)-24 \times \left| R_0 - {1 \over 2R_0}\right|.
\ee
The essential feature is a discontinuity in the first derivative of $Z$ as a function of $R_0$,
signaling a phase transition. 
The discontinuity occurs at the self-dual point
$R_0=1/\sqrt{2}$. It is sourced by the lowest mass states in the $O_8\bar V_{24}$-sector, which become
massless precisely at the fermionic point. 
Indeed, the lowest mass in the $O_8\bar V_{24}$ sector is given by
\be
m_{O\bar V}^2=\left(R_0 -{1\over {2}R_0}\right)^2,
\label{mass} 
\ee   
and it is everywhere positive except for the fermionic point, where it vanishes.
Similar behavior is exhibited in thermal configurations associated to
non-critical heterotic strings in 2d \cite{DLS}.

At this point, several comments are in order:

\noindent $\bullet$
For $R_0 > 1/\sqrt{2}$, the partition function is given by 
\be
{Z\over V_1}={24\over R_0 } \label{part+}, 
\ee
coinciding with the conventional thermal partition function of $24\times 8$ massless 
bosons and $24 \times 8$ massless fermions in two dimensions\footnote{The contribution of each 
massless boson/fermion pair to the one-loop thermal partition function in 2 dimensions is $1/(8R_0)$.}.
The corresponding temperature is $T=1/\beta$, where $\beta=2\pi R_0$. 
This result
can be obtained by Poisson re-summing over the momentum quantum number $m$, and then mapping the integral over the fundamental domain
to an integral over the strip \cite{FtoStrip},
involving the zero winding $(n=0)$ orbits\footnote{Due to the
initial $(4,0)$-supersymmetry, the contribution of the modular invariant
$(\tilde m,\, n)=(0,0)$ orbit vanishes.}. The latter are
contained in the even winding sectors $V_8\bar V_{24}$, $S_8 \bar V_{24}$, $V_8\bar S_{24}$ 
and $S_8 \bar S_{24}$. The splitting of the integral (\ref{parti}) into its orbit by orbit contributions, and the subsequent mapping to the strip, is only valid for $R_0 > 1/\sqrt{2}$, since then the individual $(\tilde m, n)\ne (0,0)$ orbit contributions in the Lagrangian lattice are exponentially suppressed as $\tau_2 \to \infty$ and we can safely exchange the orders of summation and integration\footnote{For $R_0 < 1/\sqrt{2}$, we can obtain the result either by applying $T$-duality or by
first Poisson re-summing over the winding number $n$, which effectively maps the system on a large radius. See the discussion below.}. For $R_0 = 1/ \sqrt{2}$, the lack of this absolute convergence due to the presence of additional massless states, results in the non-analytic part of (\ref{partfinal}) \cite{FKT,DLS}. So for $R_0 > 1/\sqrt{2}$ we obtain \cite{FKT}:
\be
\label{strip}
{Z \over V_1}=R_0\sum_{{\tilde m} \ne 0}^{\infty}~\int_{||} \frac{d^2 \tau}{8\pi({\rm Im}\tau)^{2}}~
e^{-{\pi({\tilde m}R_0)^2\over \tau_2}}~{\Gamma_{E_8}\over ~\eta^{8}}~
\left(V_8-(-)^{{\tilde m}}~S_8\right)
~(\bar V_{24} - \bar S_{24})~.
\ee
The sectors
$V_8\bar V_{24}$ and $S_8 \bar V_{24}$, 
which contain the initially massless bosons and
fermions, are thermally excited as in the conventional thermal deformation. 
The deformation deviates from the conventional 
one in the sectors $V_8 \bar S_{24}$ and $S_8 \bar S_{24}$. 
These two sectors are massive, with masses of the order of
the string scale, even at zero temperature. 
Due to the unbroken right-moving \emph{MSDS}
symmetry, $\bar V_{24}-\bar S_{24}=24$, and level-matching,
the contributions of all massive boson and fermion 
string oscillator states cancel. 
Only the thermally excited massless states 
give non-vanishing net contributions, leading to  
the result (\ref{part+}):
\be
{Z \over V_1}=(24\times 8)~{1 \over \pi^2R_0}~\sum_{{\tilde m} = 0}^{\infty}~{1\over \ (2\tilde m+1)^2}= {24 \over R_0}.
\ee 

\noindent$\bullet$
The integral over the strip allows one
to rewrite the partition function $Z$ in terms of (the logarithm of) 
a (spacetime) ``dressed'' thermal trace 
over the $2^{\textrm{nd}}$ quantized Hilbert space of the 
initially supersymmetric Hybrid B model. 
Because of the 
asymmetric nature of the 
deformation, this is given by the right-moving fermion index:
\be
\label{dress}
Z=\ln\textrm{tr}{\left[~e^{-\beta H}~
(-1)^{F_R}\right]}~.
\ee
This expression is strictly valid for $\beta=2\pi R_0>2\pi/\sqrt{2}$, as indicated by
the non-analyticity of the exact result for the partition function (\ref{partfinal}).
The right-moving fermion index
differs from the canonical 
thermal trace in the odd-$F_R$ sector ($\bar a =1$). 
Such states originate from the massive 
sectors $V_8 \bar S_{24}$ and $S_8 \bar S_{24}$ 
(with masses of the order of the string scale). 
As we have seen, 
in the presence of unbroken right-moving \emph{MSDS} symmetry, the
contributions of {\it all} 
states involving massive string oscillators cancel 
and so the right-moving fermion index reduces 
to a {\it canonical} thermal ensemble over 
the initially massless fields\footnote{It would be interesting to see
if these cancellations persist at the higher genus levels.}.
We can also understand the cancellations as follows. 
The contribution to $Z$ from a single bosonic (fermionic) 
field of mass $M$ and right-moving fermion number $\bar a$ is given by
\be
\mp\sum_{k}\ln \left(1\mp (-1)^{\bar a}e^{-\beta\omega_k}\right), 
\quad \omega_k=\sqrt{k^2+M^2}.
\ee
This shows that the contribution of a bosonic field with right-moving fermion number $\bar a$ cancels against 
the contribution of a fermionic field with the same mass and right-moving fermion number $1-\bar a$. 
The right-moving \emph{MSDS} symmetry requires that such massive bosons and fermions come in pairs. 

\noindent$\bullet$
Consider now the canonical ensemble, 
defined in terms of the conventional (undressed) thermal trace. 
For sufficiently large $\beta$, 
the contributions of all states involving massive string oscillators 
are suppressed 
and, effectively, the thermal configuration reduces 
to a system of thermal radiation associated 
with the initially massless states in the sectors $V_8 \bar V_{24}$ 
and $S_8 \bar V_{24}$. In this sense, the deformed model does not differ 
appreciably from the conventional thermal system for large $\beta$.  
For values of $\beta$ close to the string length, 
massive string oscillators become
relevant. The canonical ensemble fails to converge for 
$\beta < \beta_H = (2\pi) \sqrt{2}$, due to the exponential growth of the number of 
single-particle oscillator states. The introduction of
the discrete gravito-magnetic fluxes effectively regulates this behavior.
In their presence, the temperature can be further increased up to the
fermionic point $\beta = (2\pi)/\sqrt{2}$, until one encounters 
a phase transition, which, in the present model, can be followed perturbatively. 

\noindent$\bullet$
For $R_0 < 1/\sqrt{2}$, the partition function is given by
\be
{Z\over V_1}={24\times(2R_0) } \label{part-}. 
\ee
This result can be obtained by applying the $R_0\to 1/(2R_0)$ duality to
equation (\ref{part+}). Alternatively, it can be derived by Poisson-resumming over the
winding number $n$ in equation (\ref{parti}) and unfolding the integral over the fundamental domain to an integral over the strip, as before. 
Essentially, $T$-duality maps the Hybrid B system at small radius $R_0$ to a thermal Hybrid A system at a large radius $1/(2R_0)$.    
The light states occur in the $V_8\bar V_{24}$ and $C_8 \bar V_{24}$ sectors. In both of them, 
these are purely winding states (or vortices). They can be interpreted as ordinary
thermal excitations, with temperature given by the inverse period of
the $T$-dual circle: $T=1/\beta$, $\beta=2\pi/(2R_0)$. 
This definition of the temperature
in the small radius regime is in accordance with our expectation from
conventional thermodynamics: typically the one-loop thermal partition 
function decreases as the temperature decreases. We see that the Hybrid B system
at small radii is again effectively cold and better understood as a Hybrid A thermal system at large radii.

\noindent$\bullet$
At the self-dual point $R_0=1/\sqrt{2}$, 
there are $24$ complex (or $48$ real) additional
massless states arising from the $O_8\bar{V}_{24}$-sector (see Eq. (\ref{mass})). These
states carry simultaneously non-trivial momentum and winding quantum numbers along the Euclidean
time circle. Notice that, at the fermionic point, 
the corresponding values of the left- and right-moving 
momenta are given by
\be
p_L=\pm 1,\,\,\,\, p_R=0,
\label{Ocharges}
\ee
and there is a further enhancement of the local gauge symmetry
group associated with the compact Euclidean time circle \cite{FKT}:
\be
U(1)_L \times U(1)_R \to [SU(2)_L]_{k=2} \times U(1)_R~.
\ee
The effect of these states in
the partition function (\ref{partfinal}) is to induce the non-analytic part. 
The latter can also be expressed as 
$-24|m_{O\bar V}|$, which can, in turn, be understood in the following way. Near the extended symmetry
point the contribution of the lowest mass states in the $O_8\bar V_{24}$-sector is given by
\be
48 \times \int_1^{\infty}{dt \over 4\pi t}~t^{-1/2}~e^{-\pi m_{O\bar V}^2 t}
=48~ |m_{O\bar V}|~{1\over 4\sqrt{\pi}}\int_{\pi m_{O\bar V}^2}^{\infty}{dy \over y}~y^{-1/2} e^{-y}. 
\ee 
In the limit $m_{O\bar V}\to 0$, the leading $m_{O\bar V}$-dependent contribution is
\be
48~|m_{O\bar V}|~{1\over 4\sqrt{\pi}}\Gamma\left(-{1 \over 2}\right)=-24~|m_{O\bar V}|.
\ee

In summary, the system defined in Eq.(\ref{parti}) has three characteristic regimes, where different light states 
dominate the underlying effective field theory dynamics as the value
of the thermal modulus is varied. There are distinct Hybrid B and Hybrid A phases
related by ``thermal duality symmetry'',  $R_0 \to 1/(2R_0)$ together with $S_8\leftrightarrow C_8$, and a phase transition between them at the self-dual point $R_0=1/\sqrt{2}$.   
In the Hybrid B  phase $R_0 > 1/\sqrt{2}$, the light thermal
states belong to the space of momenta, whereas in the Hybrid A phase $R_0 < 1/\sqrt{2}$, the light thermal states belong to the space of windings. The third intermediate regime corresponds to the neighborhood of the fermionic point $R_0=1/\sqrt{2}$ and governs the transition between the two dual phases. 

Thus it is necessary to utilize three distinct local effective field theories in order to describe fully the stringy thermal system. Alternatively, one can use a single non-local
description in which the spacetime variables are doubled, utilizing constraint fields that depend simultaneously on variables
conjugate to the momentum and winding modes \cite{HZ}.
In our specific model, the transition between the momentum and winding spaces 
can be exactly resolved in terms of localized branes, as we will
discuss later.  

At the fermionic point, the massless states form a localized $[{SU(2)_L}]_{k=2}$
symmetric space. Operators associated with the extra massless states, which are localized 
at the fermionic point $R_0=1/\sqrt{2}$, induce transitions
between the purely momentum and winding modes. To see this,
recall that the corresponding left- and right-moving momentum charges are given
by equation (\ref{Ocharges}). We denote the corresponding vertex operators, integrated over the
worldsheet super-coordinates, $O_+$ and $O_-$
respectively. Next, consider a level-matched, purely momentum state in the $S_8\bar V_{24}$
sector. At the critical point, such states with lowest mass have $p_L=p_R=\pm {1 \over 2}$. Let us
focus on the state $|{1 \over 2},\,{1 \over 2}\rangle$.
Acting on it with $O_-$ lowers $p_L$ by one unit and leaves $p_R$ unchanged. The new state has
$p_R=-p_L={1 \over 2}$ and it is a purely winding state. It corresponds to a level-matched state 
in the $C_8\bar V_{24}$-sector.
Pictorially, the action of the $O_-$ marginal operator can be described as
\be
O_-~\left|S_8;\, {1 \over 2},\, {1\over 2}\right\rangle \rightarrow \left|C_8;\, -{1 \over 2},\, {1\over 2}\right\rangle~.
\ee  
The explicit mapping can be understood as follows. In the $0$-ghost picture, $O_{-}$ corresponds to the zero-mode of the holomorphic current $J_{-}=\psi^0 e^{-iX_L^0}$. Moreover, the $(-\frac{1}{2})$-picture vertex operator associated to the massive spacetime fermion state $\left|S_8;\,\frac{1}{2},\frac{1}{2}\right\rangle$ takes the form $e^{-\phi/2}S_{10,\alpha} e^{\frac{i}{2}X_L^0+\frac{i}{2}X_R^0}\bar{V}_{24}$, where $S_{10,\alpha}$ is the spin-field transforming in the spinorial representation of $SO(10)$ (of definite chirality) and $\bar{V}_{24}$ is the right-moving oscillator contribution from the vectorial representation of $SO(24)$. In our conventions, the spinorial representation comes with odd chirality, so that the overall vertex operator be consistent with the GSO-projection. The action of $J_{-}$ on $\left|S_8;\,\frac{1}{2},\frac{1}{2}\right\rangle$ yields:
\be
	J_{-}(z) e^{-\phi/2} S_{10,\alpha} e^{\frac{i}{2}X_L^0 + \frac{i}{2} X_R^0} \bar{V}_{24}(w) = \frac{1}{z-w}\, \gamma^0_{\alpha\dot\beta} e^{-\phi/2} C_{10,\dot\beta} e^{-\frac{i}{2}X_L^0+\frac{i}{2}X_R^0} \bar{V}_{24}(w) + \textrm{reg.},
\ee 
where $\gamma^0$ is the familiar Dirac $\gamma$-matrix in 10 dimensions. As a result of this action, the chirality of the spinorial representation changes. The new vertex operator then corresponds to a physical state in the odd winding sector of the theory.  Taking the contour integral picks up the simple pole contribution from the OPE and we readily verify that $O_{-}$ indeed maps the purely momentum state, $\left|S_8;\,\frac{1}{2},\frac{1}{2}\right\rangle$, into the purely winding state, $\left|C_8;\,-\frac{1}{2},\frac{1}{2}\right\rangle$, as stated above.

To make the duality properties of the system more transparent,  it is convenient to express the thermal modulus in terms of the variable $\sigma \in (-\infty,+\infty)$: 
\be
R_0 = {1 \over \sqrt{2}}~e^{\sigma}.
\ee
Then the \emph{physical, duality-invariant temperature} can be written neatly as follows:
\be
T= T_c ~e^{-|\sigma|}~~~~,~~~~  T_c={\sqrt{2} \over 2\pi}~.
\ee
This expression makes manifest the fact that the temperature of the system is bounded from above, never exceeding a critical value:
$T \le T_c$. As we vary the thermal modulus $\sigma$ adiabatically from $-\infty$ to $+\infty$,
the system heats up to the critical temperature; it then undergoes a stringy phase
transition and, subsequently, it cools down. 
In terms of the modulus $\sigma$ and the physical
temperature, the partition function (\ref{partfinal}) can be written as
\be
\label{Lambda}
{Z \over V_1}=(24\sqrt{2})~e^{-|\sigma|}=\Lambda~{T },\,\,\,\,~~ 
\Lambda={24\sqrt{2}\over T_c}=48\pi,
\ee    
where we introduced the thermal parameter $\Lambda$ being proportional to the
number of thermally excited light states. The two dual thermal phases occur at $\sigma >0$ and $\sigma <0$, respectively. In each phase
the pressure and energy density are given by
$$
P=\Lambda~T^2,
$$
\be
\rho=-P+T{\partial P \over \partial T}=\Lambda~T^2~.
\ee
Therefore, in each phase the equation of state is, effectively, that of thermal massless radiation in two dimensions:
$\rho =P$. At low temperature, the two dual phases are distinguished by 
the light thermally excited massless spinors: in the Hybrid B phase these transform under
the $S_8$-spinor of the internal $SO(8)$-symmetry group and in the Hybrid A
phase in terms of the conjugate $C_8$-spinor. In addition, the stringy thermal system is characterized by a critical temperature $T_c$ and, consequently, by a critical energy density given by:
\be
\rho_c=\Lambda~T_c^2=24/\pi. 
\ee

In the stringy thermal system, new light modes appear at $T_c$, which are responsible for the transition
between the two dual phases. This transition is implied by the underlying $T$-duality of the model
and the existence of an extended symmetry point. In particular, it should persist to all
orders in the string genus expansion. As we will show in the following section, this phase transition admits a
brane interpretation that glues together the space of ``momenta" and the dual space of ``windings''. 
This connection is absolutely transparent in the two-dimensional Hybrid model under consideration.
In higher dimensional thermal models, we also expect similar gluing connections between the spaces of ``momenta" and ``windings",
described via localized branes at $T_c$.  
The simple conical structure of equation (\ref{partfinal}) will be replaced by
a more involved structure according to the dimensionality of the Euclidean spacetime.


\section{Non-singular cosmology}
\label{Cosmo}
In this section, we will analyze the backreaction of the tachyon-free thermal Hybrid configuration on the
initially flat metric and dilaton background. In several models the induced cosmological
evolution in the intermediate regime $T\ll T_H$ has been extensively analyzed\cite{InducedCosmo, AtracorCosmo}, where
an interesting attractor solution corresponding to a radiation-like era was found \cite{AtracorCosmo}.  Other work in the context of hot string gas cosmology includes \cite{HotstringsCosmo}. In this work,
however, the important high temperature regime and the structure of the stringy phase transition at the Hagedorn temperature will be analyzed.    
As we have seen in section \ref{TH}, the thermal stringy system has three characteristic regimes associated to the
space of light momenta ($\sigma >0$), the space of light windings ($\sigma < 0$) and 
the extended symmetry point which is localized at $\sigma=0$ and connects the two. 
At sufficiently low temperature, the first two regimes are effectively described by 
local field theories involving the light states of the Hybrid B and A models, respectively. At the
extended symmetry point ($\sigma=0$), the effective description is in terms of a one-dimensional non-Abelian gauge 
theory, based on the  gauge symmetry group
\be
 H_L\times H_R \equiv \left(SU(2)\times U(1)^8\right)_L \times \left(U(1)\times SU(2)^8\right)_R \, ,
\ee
 in contrast to the other two regimes where $H_{L}$ is broken down to the Cartan subgroup $H_L \rightarrow (U(1)\times U(1)^8)_{L}$. 
In all regimes the massless states are scalars transforming in the (\,{\it adjoint}\,,\,{\it adjoint}\,) representation of $H_L\times H_R$.
 
The Hybrid A and Hybrid B phases are mapped into each other by thermal duality, 
implying the presence of a maximal critical temperature, $T=T_c~ e^{-|\sigma|}$. 
The non-vanishing energy density and pressure induce a cosmological evolution, during which  
the thermal modulus $\sigma$ becomes a monotonic function of time: $\sigma\rightarrow\sigma(t)$.  
Since  $\sigma$  is a monotonic function, it can be used as a parametric time variable, scanning
all three effective field theory regimes of the stringy thermal system. The cosmological evolution 
will be marked by a transition from the Hybrid A phase to the Hybrid B phase 
which, at finite coupling, is well localized at $\sigma=0$.
 The phase transition is driven by the extra thermal states that become massless at this point 
 and can be effectively described in terms of space-like
branes, localized in time. Stringy space-like
brane configurations in different context have been discussed in \cite{SG}.  
As we will see below,
the induced cosmological evolution turns out to be non-singular: thanks to the presence of the space-like branes,
the  ``Big Bang'' singularity of general relativity is absent.  

The scope of this section is to establish the existence of an effective Lorentzian action, able to describe
the three stringy regimes simultaneously, which goes beyond the leading $\alpha^{\prime}$-approximation 
and is valid up to the genus-1 level.
Our guidelines are the symmetry properties and behavior of the thermal Hybrid model and its partition function, namely:
\begin{itemize}
\item The exact expression (\ref{partfinal}), beyond any $\alpha^{\prime}$-approximation, of the 
one-loop partition function, exhibiting a specific conical
structure that is sourced by the additional massless states at $\sigma=0$.  
\item The effective equation of state in both Hybrid A and B phases,
which is that of thermal massless radiation in 2d. Thanks to the unbroken right-moving $MSDS$
symmetry, the contributions to the partition function of all massive string oscillator states cancel.
\item The thermal duality relation between the Hybrid A and Hybrid B phases, $\sigma \to -\sigma$, 
interchanging simultaneously the left-moving
spinors: $S_8 \leftrightarrow C_8$. This symmetry, together with CPT invariance (always present in an effective field theory), imply that
below the critical temperature the corresponding effective actions are related by time reversal, namely  $\sigma(t) \to -\sigma(-t)$.  
\item The existence of the extended symmetry point at $\sigma=0$ and its brane interpretation, gluing the two
phases A and B.
\end{itemize}
The above ingredients lead to an effective two-dimensional dilaton-gravity action, 
able to describe simultaneously and in a consistent way the
three regimes of the stringy Hybrid model.
The effective action is naturally  separated into three parts\footnote{In principle, higher-derivative terms have to be included in this action together with the contribution coming
from the higher genus levels.
As we will see later, such terms are suppressed during most of the cosmological evolution. 
In the region close to the phase transition their suppression is controlled by the value of
the brane tension. The existence of an
exact CFT description at the extended symmetry point allows us to have quantitative control even in this region. }:
\begin{align}
\label{action}
	\S = \S_0 + \S_1 + \S_{\textrm{brane}},
\end{align} 
where
\begin{align}
	&\S_0 = \int{d^2 x~ e^{-2\phi}~\sqrt{-g}\left(\frac{1}{2}~R+2(\nabla\phi)^2\right)},\\
	&\S_1 = \int{d^2 x~ \sqrt{-g}~ P},\\
	&\S_{\textrm{brane}} = -\kappa\int{d x^1 d\sigma~ e^{-2\phi}~\sqrt{g_{11}}~\delta(\sigma)}.
\end{align}
- $\S_0$ is the familiar genus-0 dilaton-gravity action (written in the string frame). \\
- $\S_1$ is the genus-1 contribution of the thermal effective potential $-P$.  
The absence of a dilaton factor in this term indicates its origin as a 1-loop effect.
The energy density and pressure are given by:
\be
	\rho = P = \frac{\Lambda}{\beta^2}~~~~,~~~~ \beta=\beta_c\,e^{|\sigma|}~,
\ee
where $\beta$ is the inverse proper
temperature, $\beta_c=(2\pi)/\sqrt{2}$ the
inverse critical temperature and $\Lambda=48\pi$ is the thermal parameter computed previously in Eq.(\ref{Lambda}).
We emphasize once more that our one-loop computation of the energy density and pressure
is exact in $\alpha^{\prime}$ 
as the contributions of the massive string oscillator states 
cancel due to the underlying right-moving \emph{MSDS} symmetry.\\ 
-  $\S_{\textrm{brane}}$ is the spacelike brane contribution 
at the phase transition. It gives rise to localized negative pressure and is sourced 
by the additional $24$ complex 
massless bosons, $\chi_i$ ($i=1,\dots, 24$), at the extended symmetry point $\sigma=0$.
Indeed, the microscopic origin of this term follows from the underlying description of
the system at the extended symmetry point, giving rise to a localized action:
\be
\label{ac1}
 \S_{\textrm{brane}}= \left.\int dx^1  \; \sqrt{g_{11}}\; e^{-2\phi}\left( -\; g^{11}\; {\partial \chi_i\over \partial x^1}{\partial {\bar \chi}_i\over \partial x^1}\right)\right|_{\textrm{at}~\sigma(x^0)=0}.
\ee
We will be interested in spatially homogeneous solutions for $g_{11}$ and $\phi$. 
Then the ${\bar \chi}_i$ equation of motion yields:  
 \be
\label{sol}
{\partial^2\chi_i\over {\partial x^1}^2}=0\qquad \Longrightarrow \qquad  \chi_i=\alpha_i+\gamma_i\sqrt{g_{11}}\; x^1,
\ee
where $\alpha_i$ and $\gamma_i$ are integration constants (invariant under $x^1$-reparametrization). Substituting the above solution back into the action, determines the
value and sign of the brane tension in terms of the ``rapidity" coefficients $\gamma_i$:  
\be
\label{tension}
\kappa=\sum_{i=1}^{24}\abs \gamma_i\abs^2=24\,\abs \gamma_s \abs^2.
\ee
In the most symmetric configuration all the rapidity coefficients are equal,  ($\gamma_i = \gamma_s,\, i=1,...,24$), and so
the brane tension is proportional to the number of the extra massless states.

Generically, in the effective action, several other light fields appear; namely, there are in total $64$ moduli fields parametrizing
the coset manifold 
\be \label{manifM}
 {\rm\cal M}={SO(8,8)\over SO(8) \times SO(8)}~.
\ee
 In all field theory phases of the thermal Hybrid model, the one-loop effective potential lifts these flat directions \cite{AtracorCosmo,StabilizationThermal}
 and the above moduli are attracted to the enhanced \emph{MSDS} symmetry points. This justifies our inclusion in the 
 action of only the dilaton-gravity background fields and the contribution of the brane. Thus, in all subsequent analysis, we will work in the case where all remaining dynamical (transverse) moduli are frozen to their \emph{MSDS} values so that the right-moving \emph{MSDS} symmetry of the theory is preserved. 
   
 We are now in a position to determine the induced stringy cosmological evolution by deriving the
 equations of motion in the presence of the brane term. 
We parametrize the metric as follows ($x^0\equiv t$):
\be
	ds^2 = -N^2\,dt^2+a^2\, {dx^1}^2,
\ee
where the lapse function $N$ and scale factor $a=e^{\lambda}$ are functions of the time coordinate only. 
Without loss of generality, we choose the phase transition at $\sigma=0$ to occur at $t=0$. 
In terms of these variables, the action, valid for all times, is given by:
\be
	\S =\int{dt\, d x~e^{-2\phi+\lambda}~\left(\frac{1}{N}\left[ 2\dot{\phi}\dot{\lambda}-2\dot\phi^2\right]-\kappa\delta(t)\right)}+\int{dt\,dx~N e^{\lambda}~P}\, .
\ee
Varying with respect to $N$, $\phi$, $\lambda$, we obtain the following system of equations:
\begin{align}
	&\dot\phi^2-\dot\phi\dot\lambda=\frac{1}{2}N^2 e^{2\phi}\rho,\\
	&\ddot\lambda-\dot\lambda\left(2\dot\phi-\dot\lambda+\frac{\dot N}{N}\right) = N^2 e^{2\phi} P,\\ \label{tria}
	&\ddot{\phi}-2\dot\phi^2+\dot\phi\left(\dot\lambda-\frac{\dot N}{N}\right)=\frac{1}{2} N^2 e^{2\phi}(P-\rho) -\frac{1}{2}N \kappa \delta(t)=-\frac{1}{2}N \kappa \delta(t).
\end{align}
In deriving the first equation, we use the thermodynamical identity\footnote{This equation exemplifies a deep connection between gravity and thermodynamics,
since in the Euclidean description, where $N= e^{\abs \sigma \abs}$ (with $x^0\sim x^0+2\pi$), the thermodynamical identity is simply derived by
the Euclidean gravity equations of motion. This non-trivial connection of gravity and thermodynamics will be explored in future work.} 
\be
-P-N{\partial P \over \partial N}=-P-\beta{\partial P \over \partial \beta}=\rho,
\ee
following from the fact that in any frame the inverse proper temperature, or the period of
the Euclidean time direction, is proportional to the lapse
function $N$. 
The r.h.s. of equation (\ref{tria}) was simplified further by using the equation of state $\rho=P$. 
The crucial effect of the brane is to induce a discontinuity in the
first time derivative of the dilaton at $t=0$.

The thermal entropy (per co-moving unit cell) is given by 
\be
\label{Entropy}
S=a\beta(\rho+P)= 2\Lambda\,\frac{a}{\beta}.
\ee
On each side of the brane this quantity is conserved, as can be derived by
using the equations of motion. At the brane, the temperature  is fixed attaining its
maximal critical value, $\beta=\beta_c$. Continuity in the scale factor across the
brane implies the continuity of the entropy and vice-versa.  
Equivalently, the stringy phase transition does not produce latent heat and is similar to
a second order transition.
More precisely, by utilizing the equations of motion, we 
derive the following equation:
\be\label{entropyCons}
	\dot{\rho}+\dot\lambda(\rho+P)= \frac{2\kappa}{N} e^{-2\phi} \left(\dot{\lambda}-2\dot{\phi}\right)~\delta(t),
\ee
where the r.h.s. represents the brane contribution.
Integrating (\ref{entropyCons}), we find that the thermal entropy is given by
\be
	S(t)=S(t_0)+ \kappa\beta_{\textrm{c}}\frac{a(0)}{N(0)} 
e^{-2\phi(0)}\left[\dot\lambda(0^+)+\dot\lambda(0^-)-2\left(\dot\phi(0^+)+\dot\phi(0^-)\right)\right]\vartheta(t),
\ee
where $\vartheta(t)$ is the step function and $t_0<0$.
Continuity of the thermal entropy\footnote{Some generalized solutions will be discussed in Section \ref{General}.} then imposes 
the following condition for the discontinuities in the 
first time derivatives of the scale factor $\lambda$ and the dilaton $\phi$:
\be\label{EntropyCondition}
	\dot\lambda_{+}+\dot\lambda_{-}=2\left(\dot\phi_{+}+\dot\phi_{-}\right),
\ee 
where from now on we simplify the notation to $\dot\lambda(0^{\pm})=\dot\lambda_{\pm}$, 
$\dot\phi(0^{\pm})=\dot\phi_{\pm}$. 
When the thermal entropy is conserved, we obtain the following relation between the scale factor and the thermal modulus $\sigma$:
\be\label{relation1}
\lambda = |\sigma|+\ln{\frac{\beta_{\textrm{c}}S}{2\Lambda}}.
\ee
Therefore, the scale factor $a$ at any time must be greater than (or equal to) 
a certain critical size
\be
a \ge \beta_c ~{S \over 2 \Lambda}, 
\ee 
hinting at the absence of the conventional Big Bang singularity of general relativity. 

The equations of motion simplify considerably by choosing an appropriate gauge. 
To this end, we utilize time-reparametrization invariance and work in the conformal gauge $N=a$. 
The equations of motion then become:
\begin{align}\label{one}
	\textrm{(i)}~~:~~&{2}\dot\phi^2-{2}\dot\phi\dot\lambda={C} e^{2\phi},\\ \label{two}
	\textrm{(ii)}~~:~~&\ddot\lambda-{2}\dot\phi\dot\lambda = C e^{2\phi} ,\\ \label{three}
	\textrm{(iii)}~~:~~&{2}\ddot{\phi}-4\dot\phi^2= - \kappa e^{\lambda_0} \delta(t),
\end{align}
where $a_0 \equiv e^{\lambda_0} \equiv e^{\lambda(0)}={\beta_{\textrm{c}}S /2\Lambda}$ is the value of the scale factor 
when the system reaches its critical temperature and
\be
\label{C}
	C \equiv \frac{S^2}{4\Lambda}
\ee
is a constant being proportional to the number of light degrees of freedom.

Subtracting Eq.(\ref{one}) from (\ref{two}), we obtain $\ddot{\lambda}-2\dot{\phi}^2=0$. 
Further subtracting the latter from Eq.(\ref{three}), we obtain 
$2\ddot{\phi}-2\dot{\phi}^2-\,\ddot{\lambda}= -\kappa e^{\lambda_0}\, \delta(t)$. 
Observe that both of these equations as well as Eq.(\ref{three}) are independent of the constant $C$. Their structure 
clearly shows that only the first time derivative of the dilaton is discontinuous across the phase transition, while its 
second derivative gives rise to the delta function brane source. 
The first time derivative of the scale factor is continuous across the brane: $\dot\lambda_{+}=\dot\lambda_{-}$. 
The discontinuity in $\dot{\phi}$ is given in terms of the brane tension:
\be\label{relation2}
	\dot\phi_{+}-\dot\phi_{-} = - \frac{1}{2}\, e^{\lambda_0} \kappa.
\ee
Using condition (\ref{EntropyCondition}) together with the continuity of $\dot\lambda$, 
we obtain $\dot\lambda(0)=\dot\phi_{+}+\dot\phi_{-}$.
 The conservation of the entropy (\ref{Entropy}) then implies
\be
	\dot{\lambda} = {\dot{\beta} \over \beta}~,	
\ee
showing that the first time derivative of the temperature is also continuous across the critical point. 
The temperature attains its maximal (critical) value $T_{\textrm{c}}$ at $t=0$ and, thus, 
its derivative and the derivative of the scale factor must both vanish at this point: 
$\dot{\beta}(0)=\dot\lambda(0)=0$. 
Eq.(\ref{EntropyCondition}) then implies that the first time derivative of the dilaton 
flips sign $\dot\phi_{+}=-\dot\phi_{-}$ across the brane. The net result 
shows that the dilaton bounces during the transition. 

Solving Eq.(\ref{three}) we obtain the following expression for the time dependent dilaton:
\be\label{DilatonSolution}
	e^{2\phi(t)} = {e^{2\phi_0} \over {1+2\dot{\phi}_{-}|t|  } }~\equiv~g^2_{\rm str}(t)\, ,
\ee
where $ \dot\phi_{-}$ is a positive constant determined in terms of the brane tension $\kappa$
\be\label{relation3}
	 \dot\phi_{-} =-\dot\phi_{+} =  \frac{1}{4}\, e^{\lambda_0} \kappa\, .
\ee
The above expression shows that the string coupling $g^2_{\rm str}(t)$ remains smaller than a maximal value, 
$g^2_{\rm str}(0)=e^{2\phi_0}$, for all times and, therefore,
the perturbative validity of the model is guaranteed, provided that $g^2_{\rm str}(0)$ is
sufficiently small. The dilaton reaches its maximal value at the critical point and then decreases monotonically. 

It turns out that the string coupling $g_{\rm str}(0)$ is not an arbitrary parameter. Indeed, from Eq.(\ref{one}) 
evaluated at the critical point we obtain a relation between $C$ and the derivative $\dot\phi_{-}$ of the dilaton 
\be
	\dot\phi_{-}^2 = \frac{C}{2}\, e^{2\phi_0}.
\ee
Combined with (\ref{relation3}), it yields:
\be
	e^{2\phi_0} = \frac{\kappa^2 \beta_{c}^2}{8\Lambda} = 3\pi \abs \gamma_s \abs^4,
\ee
where the last expression in terms of the rapidity coefficient $\gamma_s$ is obtained by utilizing Eq.(\ref{tension}), 
relating the brane tension $\kappa$ with the rapidity, the value of the critical temperature $\beta_{\textrm{c}}^{-1}$ 
and the numerical value of the thermal parameter $\Lambda$. The validity of the perturbative regime is
ensured by the smallness of the rapidity coefficient $\gamma_s$. 

The evolution of the scale factor is obtained by solving Eq.(\ref{two}). The result is:
\be\label{ScaleFactorSolution}
	a^2(t)=e^{2\lambda(t)} = e^{2\lambda_0}{ e^{2\dot\phi_{-}|t|} \over 1+2\dot{\phi}_{-}|t| }.
\ee
It is convenient to perform a coordinate transformation $\tau=2\dot\phi_{-}t$, $x=2\dot\phi_{-}x^1$. In the new coordinates, the
cosmological solution for the metric and dilaton is:
\begin{align}
	& ds^2 = \frac{4}{\kappa^2}~\frac{e^{|\tau|}}{1+|\tau|}~\left(\, -{d\tau}^2 + {dx}^2 \, \right)\, ,\\
	& ~g^2_{\rm str}(\tau) \equiv e^{2\phi(\tau)} = \frac{ e^{2\phi_0} }{1+|\tau|}~ =~{\pi \kappa^2\over 192}~ {1 \over 1+|\tau|}~ \, .
\end{align}
The metric has no essential singularities; both the Ricci scalar 
\be
	R = \frac{\kappa^2}{4}~ \frac{e^{-|\tau|}}{1+|\tau|},
\ee
and the string coupling are bounded from above. Their maximal values occur at the critical point 
and are both proportional to the square of the brane tension,
$\mathcal{O}(\kappa^2)$.
As a result, all higher-derivative terms, as well as higher-genus corrections, 
are controllable by a perturbative expansion in terms of $\kappa$, provided that $\kappa=24|\gamma_s|^2 \ll 1 $. 

The above solution, even for large times, differs from the naive radiation-dominated cosmology,
mainly because of the presence of a running dilaton. To see this, we switch to the cosmological
frame, where the metric takes the form
\be
	ds^2 = -d\xi^2 + a^2(\xi)d\tilde x^2.
\ee
The cosmological time $\xi$ is given in terms of the conformal parametric time $\tau$ as:
\be
	\xi(\tau) = \frac{2|\tau|}{\kappa} \int\limits_0^1{\frac{e^{|\tau| u/2}}{\sqrt{1+|\tau| u}}~du} 
	= \sqrt{\frac{8\pi}{e\kappa^2}}\left[ \textrm{erfi}\left(\sqrt{\frac{1+|\tau|}{2}}\right)-\textrm{erfi}\left(\frac{1}{\sqrt{2}}\right)\right],
\ee
with $\textrm{erfi}(x)\equiv \frac{1}{i}\,\textrm{erf}(ix)$ being the imaginary error function and $\tilde{x}= x/2$. 
For large cosmological times $\xi$, the asymptotic properties of the error function imply:
\be
	\xi = \frac{4}{\kappa} \frac{e^{|\tau|/2}}{\sqrt{|\tau|}}\left[1+\mathcal{O}\left(\frac{1}{ |\tau|}\right)\right]~~~~\Longrightarrow
	~~~~e^{\abs \tau\abs} = \left({\kappa \xi\over 4}\right)^2~\ln(\kappa\xi)^2 + \cdots 
\ee
In the very late cosmological time regime, the asymptotic expansions for the scale factor and string coupling are given by
\be
	a(\xi) =  |\xi| \left(1-\frac{1}{\ln{(\kappa \xi)^2}}+\cdots\right)~,~~~{1 \over g_{\rm str}^2}=
	{192 \over \pi \kappa^2} \left(\,\ln(\kappa\xi)^2  + \ln \ln (\kappa\xi)^2  + \rm{const.} + \cdots\, \right)\,.
\ee
For large $\abs \kappa \xi \abs\gg 1$, the cosmology asymptotes to the flat thermal Milne Universe. There are
logarithmic corrections due to the running dilaton, giving rise to the above behavior. 
For small cosmological times $\abs \kappa \xi \abs\ll 1$, the scale factor and the string coupling have the following expansion:
\be
	a(\xi) = \frac{4}{\kappa}\left(1+ \frac{1}{16}(\kappa \xi)^2 + \mathcal{O}(|\kappa \xi|^3) \right),~~~~
	{1 \over g_{\rm str}^2}={192 \over \pi \kappa^2}\left[ 1 + {\abs \kappa \xi \abs \over 2} - {1 \over 8} (\kappa \xi)^2 + \mathcal{O}(|\kappa \xi|^3) \right]\,,  
\ee
illustrating very clearly that the cosmological scale factor bounces at the location of the brane. The dilaton, on the 
other hand, exhibits a conical structure, which results in a discontinuity in its first time derivative.  

\section{Generalized solutions} \label{General}

There are two linear combinations of the equations of motion which are {\it independent} of
the genus-1 thermal contribution, namely
\be
\label{philambda}  
{2}\ddot{\phi}-4\dot\phi^2= - \kappa e^{\lambda_0} \delta(t)~~~~, ~~~~~
2\ddot{\phi}-2\ddot{\lambda}=-\kappa e^{\lambda_0}\, \delta(t)\;.
\ee
The third equation
$$
{2}\dot\phi^2-{2}\dot\phi\dot\lambda={C} e^{2\phi},
$$
depends on the thermal contribution through the conserved quantity $C$, which is determined by the thermal entropy; 
see Eq.(\ref{C}). 
The most general solution of the first two $C$-independent equations is the following:
 \be
 e^{2\phi}={ e^{2\phi_0}\over 1+A\abs t\abs +Bt }~,~~~~~ e^{2\lambda}=e^{2\lambda_0}\,{ e^{A\abs t\abs +\Gamma t } \over 1+A\abs t\abs +Bt }~~~~~~~{\rm with }~~~~~~A={\kappa \over 2}\, e^{\lambda_0}~.
 \ee 
It is very important that the integration constant $A$, multiplying the conical part $\abs t\abs$, 
is fixed by the brane tension $\kappa$ and is, therefore, positive. 
All solutions satisfying $\abs B \abs \le A$ avoid the infinite coupling constant singularity and 
string perturbation theory 
remains valid throughout the cosmological evolution. 
Substituting the general solution to the third equation, yields the
following relations among the integration constants:
\begin{equation}
\label{condi}
\begin{array}{l}
\dis (A+B)(A+\Gamma)=2C_+ e^{2\phi_0}, ~~~{\rm for} ~~~t>0\, ,\\\\
\dis (A-B)(A -\Gamma)=2C_- e^{2\phi_0},~~~{\rm for} ~~~t<0\, ,
\end{array}
\ee
where we allow for the possibility of having a change in thermal entropy across the brane at $t=0$, giving rise to different 
conserved quantities
$C_+$ and $C_-$ at the two sides. 
Continuity of the thermal entropy, $C_+=C_-=C\ge 0$, imposes $B+\Gamma=0$. In this case the
weak coupling constraint $\abs B \abs \le A$ is automatically satisfied. 
The restriction of not exceeding the maximal temperature at
the critical point, which is equivalent to $\dot\lambda(0)=0$, further imposes $B=\Gamma=0$ and  we obtain
the non-singular, time-reversal invariant solution of section \ref{Cosmo}. 

We proceed to explore other solutions with $C_+\ne C_-$, preserving, however, the maximal temperature
constraint, which requires $B=\Gamma$. A very intriguing non-singular solution arises when $C_-=0$, which also saturates the 
weak coupling constraint, $A=B=\Gamma$:  
 \be
 \label{thermal+}
 e^{2\phi}={ e^{2\phi_0}\over 1+A\,[\abs t\abs +t] }~~~~,~~~~~~ e^{2\lambda}=e^{2\lambda_0}\,{ e^{A\,[\abs t\abs + t] } \over 1+A\,[\abs t\abs +t ]} \, .
 \ee 
For $t<0$, the spacetime metric is flat, $\lambda=\rm const.$, and the dilaton is constant. 
For $t>0$, the Universe is filled with thermal radiation. 
The physical relevance of this solution can be understood as follows. The flat $t\le 0$ region, including the brane at $t=0$, admits a Euclidean interpretation. The Euclidean
path integral over it can be used to define the wave-function for the cosmology at $t>0$, 
which summarizes the very early-time history of the Universe, while for later times the Euclidean brane creates the radiation.
  
To understand this, we consider the Euclidean solutions, whose integration constants satisfy Eq.(\ref{condi}) after flipping the signs of the right hand sides. For $C_-=0$, $C_+>0$, the solution maintaining $A=B$ satisfies $2A(A+\Gamma)=-2C_+e^{2\phi_0}= -(2A)^2$. The last equality follows by identifying the brane tensions in the Euclidean and Lorentizian solutions and yields $\Gamma=-3A$. In half of this solution, for Euclidean time $t_E=z<0$, the dilaton is constant and the metric takes
the form:
 \be
 \label{Eflat}
 ds^2=e^{2\lambda_0-4Az}\left[ dz^2+dx^2\right] =d\rho^2+\rho^2dy^2\qquad \where\qquad 2A\rho=e^{\lambda_0-Az}~,~~~y=2Ax~.
 \ee  
In terms of the polar-like coordinates $\rho$ and $y$, the metric is locally flat, with the radial coordinate 
$\rho$ being bounded from below: $\rho\ge e^{\lambda_0}/(2A)$. 
When $y \to i\omega$, the metric takes the static Rindler form, where $\omega$ plays the role of time. Although this space
is locally flat, it is also intrinsically thermal possessing an effective ``geometrical" (entanglement) entropy. Thus, even though $C_-$ is zero in this system,
we can associate to the half-solution a non-trivial entropy. The Rindler geometrical entropy is transformed via the brane
into the thermal entropy of the late-time cosmological evolution. The half-Euclidean solution $z\in(-\infty,\, 0]$, which contains
the brane, can be used to define the wave-function of the Universe
described by Eq.(\ref{thermal+}) at $t=0$. This mechanism will be investigated further in future work.
  
  Another interesting Euclidean solution, with $C_{+}=C_{-}=0$, corresponds to $A=B=-\Gamma$, 
including the brane localized at Euclidean time $t_E=z=0$:
\be
 e^{2\phi}={ e^{2\phi_0}\over 1+A\,[\abs z \abs +z ]}~~~~,~~~~~~ e^{2\lambda}=e^{2\lambda_0}\,{ e^{A\,[\abs z  \abs - z] } \over 1+A\,[\abs z \abs +z]} \, .
\ee
As before, when $t_E=z<0$, the dilaton is constant and the metric is flat, taking
the form given in Eq.(\ref{Eflat}) under the replacement $A\to A/2$. On the other side of the brane, $t_E=z>0$, both the metric and the dilaton become non-trivial
  \be
 e^{2\phi}={ e^{2\phi_0}\over 1+2A\,z }~,~~~~~~ ds^2=e^{2\lambda_0}~{ dz^2+dx^2 \over 1+2A\,z}~.
 \ee
After an appropriate coordinate change, this solution becomes:
\be
 \label{Edualflat}
e^{2\phi}= {e^{2\phi_0+2\lambda_0}\over A^2}  {1\over \rho^2}\; , \quad ds^2=d\rho^2+{1\over \rho^2} \, dy^2\quad \where\quad 
A\rho=e^{\lambda_0}\sqrt{1+2Az}\; ,\quad y={e^{2\lambda_0}\over A}\,x ~.
\ee
In this parametrization, we recognize the well-known $T$-dual geometry of the two-dimensional flat space. Here, again, the coordinate 
$\rho$ is bounded from below, $\rho\ge e^{\lambda_0}/A$, thus avoiding the well-known singularity of the dual-to-flat metric at 
$\rho=0$. The string coupling is bounded from above by its value at the position of the brane, $g_{\rm str}^2 \le e^{2\phi_0}$. 
The dual-to-Rindler space is obtained by $y \to i\omega$, and so in this sense, it is also thermal possessing a non-zero
``geometrical" entropy.
On either side of the brane, the Euclidean solution is supersymmetric \cite{KKL}, 
 as  is also implied by the choice $C_{\pm}=0$. The brane, however, breaks supersymmetry locally and creates the
 geometrical entropy in the Euclidean. 
 The Euclidean solution clarifies further the
importance of the brane. Namely, the brane glues the two dual geometries together in a consistent way.

The time-reversed solution of Eq.(\ref{thermal+}) has $C_+=0$ and $C_-=C$; it can be obtained when 
$B=-A<0$. In this case the spacetime is flat and the dilaton is constant for positive times $t>0$.
The Universe is filled with thermal radiation at negative times. As a result, the Universe is static with 
a thermal Rindler interpretation at later times.
In this cosmological solution, 
the thermal entropy at early times is transfomed via the brane at $t=0$ to the geometrical entropy associated
to the late-times Rindler space.  

Another interesting solution is obtained by  
combining these two solutions, assuming  branes localized at two different times, namely at $t=\pm \alpha$. 
A time-reversal invariant solution can be easily derived, 
\be
 e^{2\phi}={ e^{2\phi_0}\over 1+A\,[\abs t+\alpha \abs +\abs t-\alpha \abs-2\alpha] }~~,~~~~~~
 e^{2\lambda}=e^{2\lambda_0}\,{ e^{A\,[\abs t+\alpha \abs +\abs t-\alpha \abs-2\alpha ] } \over 1+A\,[\abs t+\alpha \abs +\abs t-\alpha \abs-2\alpha] } ~~ ,
\ee
with  the following behavior:\\
i) for $t<-\alpha$, the solution for the dilaton and the metric is non-trivial, describing a contracting thermal Universe, with entropy coefficient  $C e^{2\phi_0}=2A^2$.\\
ii) for $t>\alpha$, the dilaton and the metric have non-trivial time-dependence, 
describing an expanding thermal Universe with the same entropy coefficient as in i), $C e^{2\phi_0}=2A^2$.\\
iii) in the intermediate region $-\alpha<t<\alpha$, the metric looks locally flat and the dilaton is constant.

The intermediate flat region $t \in [-\alpha,\, \alpha]$, including the branes, admits a Euclidean interpretation,
and as we explained before in the half-solutions it possesses a geometrical entropy and temperature in the
Rindler cosmological frame. Thus the thermal entropy of region i) is transformed via the
first brane at $t=-\alpha$ into the geometrical entropy of the intermediate region, which in turn is transformed back to
thermal entropy in region ii) via the second brane at $t=\alpha$. 
In the limit $2\alpha \to 0$, we recover the 
solution presented in section \ref{Cosmo}. Thus, the parameter $\alpha$, describing the separation of the branes, provides
a new relaxation time scale for the contracting-to-expanding cosmological phase transition. 
The whole picture can be understood via
Euclidean instanton transitions between contracting and
expanding thermal Universes. The pure Euclidean as well as the Rindler and
dual-to-Rindler solutions will be investigated further in future work. 

\section{Conclusions}

In this work we identified string-theoretic ingredients which can be used to resolve both 
the Hagedorn instabilities of thermal string theory, as well as 
the initial ``Big Bang'' singularity of the induced cosmological evolution. 
In order to incorporate these ingredients,
it was necessary to work beyond the low energy effective field theory approximation, 
examining purely stringy phenomena. 
For technical reasons, we focused on simple $(4,0)$-supersymmetric string models, where  
adequate quantitative control over the dynamics is available.
More specifically, we considered the class of two-dimensional Hybrid models, which are
characterized by a very special asymmetry between the left- and right-moving sectors. 
As in the usual superstring models, 
the left-moving sector is supersymmetric, giving rise to the $16$ real spacetime supercharges. 
The right-moving sector   
is non-supersymmetric at the massless level, but it enjoys the \emph{MSDS} symmetry structure, which
guarantees boson/fermion degeneracy in all of the right-moving massive levels.

We considered tachyon-free, thermal configurations of the Hybrid models, where in addition 
to the temperature, we turn on certain discrete
``gravito-magnetic'' fluxes threading the compact Euclidean time circle. 
These fluxes inject non-trivial momentum and winding charges 
into the thermal vacuum, which effectively regulate the contributions to the thermal partition function of the    
massive string oscillator states and restore the thermal $T$-duality symmetry 
of the finite temperature stringy system. 
As we have seen, the tachyon-free thermal configurations can be described in terms of freely acting asymmetric orbifolds, 
which act thermally on the left-movers but leave the right-moving \emph{MSDS} 
symmetry unbroken. Thanks to the latter symmetry, the one-loop thermal partition function was explicitly   
calculated, beyond any $\alpha^{\prime}$-approximation. 

The thermal Hybrid systems have three characteristic regimes, associated to the space of light thermal momenta, 
the space of light thermal windings and the extended symmetry point, which arises 
at the self-dual value of the thermal modulus and connects the two. 
At the extended symmetry point, extra massless thermal states
appear, with a clear brane interpretation in the Euclidean. 
These localized states are responsible for a phase transition 
between the two
dual thermal phases. The physical temperature of the system is bounded from above,
attaining its maximal critical value at the phase transition. 

An important result of this work is that we succeeded in writing down a stringy effective Lorentzian action 
able to incorporate the three stringy regimes simultaneously, which is exact in $\alpha^{\prime}$ at the two derivatives level and is valid up to genus-1.
The essential feature in this action is the spacelike brane contribution that glues together 
the spaces of light thermal windings and light thermal momenta. 
This brane is sourced by the additional massless thermal states, which appear at the extended symmetry point. 
We explicitly showed how the sign and the value of the brane tension are determined in terms
of the ``rapidity'' coefficients characterizing 
the classical backgrounds of the corresponding localized thermal fields. 
The spacelike brane, together with the bulk thermal corrections, induce
a {\it non-singular} cosmological evolution, describing a bouncing thermal Universe. 
The bounce coincides with the phase transition between the two dual Hybrid thermal phases.
The cosmology remains in the weak coupling and low-curvature regimes, provided that the rapidity coefficients
which determine the tension of the brane are small. Continuity of the thermal entropy across the brane
uniquely determines the cosmological solution, which is time-reversal invariant. 

Separated spacelike branes, localized at different times, transform the thermal entropy
into a geometrical entropy associated with a quasi-static thermal Rindler space (or
its dual) and vise-versa. Utilizing this property of the branes, we
obtained more general, time-reversal invariant solutions, which are also free
of the initial Big Bang singularity.  
The obtained non-singular solutions  
describe transitions between a contracting thermal Universe 
and an expanding one. The effect of the spacelike branes is to induce an intermediate relaxation-time period during which  
the scale factor of the Universe is quasi-static, and the entropy appears in
a geometrical Rindler-like form. The intermediate locally flat region, 
including the branes, admits a Euclidean interpretation describing
instanton transitions among contracting and expanding thermal Universes. We would like to investigate
this connection further in the near future.  

To our knowledge, the two-dimensional thermal Hybrid model presented in this work is the first example in the literature where both the stringy Hagedorn singularities as well as the classical Big Bang singularity are both successfully resolved.  

We believe that phenomenologically relevant, higher dimensional string models exist, where 
the essential ingredients, used in the thermal Hybrid model to overcome the initial singularity problem,
are also present. Namely:
i) the restoration of thermal duality symmetry in the presence of non-trivial discrete 
``gravito-magnetic'' fluxes, 
ii) the appearance of new light thermal states at the self-dual point, giving rise to a 
localized higher-dimensional brane interpretation and
iii) the gluing of dual thermal phases via higher dimensional spacelike branes, 
before the occurrence of the Big Bang singularity that is always present in the naive field theory approximation.
We conjecture that in the higher dimensional cases, the resulting duality structure
which encompasses all these ingredients will be much more intricate, 
involving other supersymmetry breaking scales and perhaps non-perturbative string phenomena as well.  
Although the technical challenges in the higher dimensional cases are expected to be more delicate, the key property, which is the joint effect of string dualities and branes ``protecting'' the stringy
system from the occurrence of spacelike cosmological singularities, could still be realized. 
Obviously, more work needs to be dedicated to this ambitious direction in order to obtain
non-singular cosmological models that will be phenomenologically viable.

  
 \section*{Acknowledgements}
 
 We are grateful to C.~Bachas, M.~Bianchi, R.~Brandenberger, C.~Callan, J.~Estes, M.~Green, W.~Lerche, D.~Luest, S.~Patil, T.~Tomaras and J.~Troost for fruitful discussions. 
I.F., C.K. and H.P. would like to thank the University of Cyprus for hospitality.
N.T. and H.P. acknowledge the Laboratoire de Physique Th\'eorique of Ecole Normale Sup\'erieure for hospitality.
C.K and H.P. thank C.E.R.N Theory Division where part of this work was discussed.   
N.T. would like to thank the Centre de Physique Th\'eorique of Ecole Polytechnique for hospitality. 
 The work of I.F., C.K. and H.P. is partially supported by the ANR 05-BLAN-NT09-573739,  and a PEPS contract. The work of I.F., C.K., H.P. and N.T. is also supported by the CEFIPRA/IFCPAR 4104-2 project. The work of H.P. is partially supported by the EU contracts PITN GA-2009-237920 and ERC-AG-226371, and PICS contracts France/Greece 
 and France/USA.\\

\end{document}